\documentclass[pre,twocolumn]{revtex4}
\usepackage{amssymb,amsmath,amsthm}
\usepackage[dvips]{graphicx}
\usepackage{verbatim}

\begin{document}
\title{CONTINUOUS-TIME QUANTUM WALKS AND TRAPPING}
\author{ELENA AGLIARI $^{*,\dag}$,
OLIVER M\"{U}LKEN $^{*}$ and ALEXANDER BLUMEN $^{*}$}
\affiliation{$^*$ Theoretische Polymerphysik, Universit\"{a}t Freiburg,
Hermann-Herder-Str. 3, D-79104 Freiburg, Germany.\\
$^\dag$ Dipartimento di Fisica, Universit\`{a} degli Studi di Parma, viale Usberti 7/A, 43100 Parma, Italy.}

\begin{abstract}
Recent findings suggest that processes such as the electronic energy transfer through the photosynthetic antenna display quantal features, aspects known from the dynamics of charge carriers along polymer backbones. Hence, in modeling energy transfer one has to leave the classical, master-equation-type formalism and advance towards an increasingly quantum-mechanical picture, while still retaining a local description of the complex network of molecules involved in the transport, say through a tight-binding approach.

Interestingly, the continuous time random walk (CTRW) picture, widely employed in describing transport in random environments, can be mathematically reformulated to yield a quantum-mechanical Hamiltonian of tight-binding type; the procedure uses the mathematical analogies between time-evolution operators in statistical and in quantum mechanics: The result are continuous-time quantum walks (CTQWs). However, beyond these formal
analogies, CTRWs and CTQWs display vastly different physical properties. In particular, here we
focus on trapping processes on a ring and show, both analytically and numerically, that distinct configurations of traps (ranging from periodical to
random) yield strongly different behaviours for the quantal mean survival probability, while classically (under ordered conditions) we always find an exponential decay at long times.

\bigskip

\noindent {\it Keywords:} Quantum Walks; Random Walks; Exciton transport; Trapping; Perturbation theory.

\end{abstract}

\maketitle

\section{Introduction} \label{sec:intro}
The transport properties of excitons in organic as well as in inorganic
molecular solids are of fundamental interest [Kempe, 2003;  Woerner {\it
et al.}, 2004; M\"{u}lken {\it et al.}, 2006; Sillanp\"{a}\"{a} {\it et
al.}, 2007; Olaya-Castro {\it et al.}, 2008; Zhou {\it et al.}, 2008].  In
general, at high temperatures the transport is incoherent and can be
efficiently modeled by continuous-time random walks (CTRWs) over sets of
participating centers (atoms, molecules, etc.) [Montroll \&
Weiss, 1965].
In this case the transport follows a master equation.
The transfer rates between the participating centers can be related to the
spatial arrangement of the centers. The arrangement is captured by the
so-called Laplacian Matrix  $\mathbf{L}$ which we will
identify here with the transfer matrix of the CTRW.
However, when dealing with
quantum particles at low densities and low temperatures, decoherence can
be suppressed to a large extent: The study of transport in this regime
requires different modeling tools, able to mimic the coherent features.
Clearly, quantum mechanical transport phenomena follow Schr\"{o}dinger's
equation. In order to make contact to CTRW
we relate the Hamiltonian of the system to the
classical transfer matrix, $\mathbf{H}=-\mathbf{L}$; this yields a
description mathematically closely connected to the classical master
equation approach.  Indeed, this realizes a quantum mechanical analog of
the CTRWs defined on a discrete structure, i.e. the so-called
continuous-time quantum walks (CTQWs). However, apart from formal
analogies, coherence can give rise to very peculiar properties (e.g.
Anderson localization [Anderson, 1958], crucial dependences of the
transport on the starting position [M\"{u}lken {\it et al.}, 2006] and a
quadratic speed-up of the chemical distance covered [Agliari {\it et al.},
2008]) with no counterpart in classical transport. These effects allow interesting and
cross-disciplinary applications and can also be exploited in experiments
in order to distinguish whether the transport is rather classical or
rather quantum mechanical [M\"{u}lken {\it et al.}, 2007].

In particular, a common means for probing the transport relies on the interaction with other species, such as impurity atoms or molecules (found in or doped into the medium) which irreversibly trap the charges or quench the excitations. Consequently, a great deal of recent theoretical work has focused on investigating essential features of basic trapping models, wherein a single particle moves in a medium containing different arrangements of traps. Indeed, much is known about the decay when the motion is incoherent [Van Kampen, 1981; Blumen {\it
et al.}, 1983; 1986; ben-Avraham \& Havlin, 2000], while (as we will show here) when quantum effects become important, strong deviations from the classical results occur.

In a set of early works the dynamics of coherent excitations on a chain with randomly distributed traps has been investigated using several methods [Hemenger \textit{et al}., 1974; Kenkre, 1978; Huber, 1980; Parris, 1991] which provided a reasonably description of the process at short times and in the asymptotic limit. On the other hand, from the experimental point of view, the most relevant regime is the one of intermediate times; in this time interval some striking effects have been recently highlighted [M\"{u}lken {\it et al.}, 2007; M\"{u}lken {\it et al.}, 2008].

Here we focus on trapping processes taking place on a finite ring where the traps are distributed according to different arrangements: the traps are either gathered in a cluster or distributed periodically or randomly. In these cases the classical survival probability $P_M(t)$ has been studied intensively (see e.g. [ben-Avraham \& Havlin 2000]). In fact, under ordered conditions $P_M(t)$
is known to exponentially decay to zero. Conversely, for a random distribution of traps $P_M(t)$ exhibits different time regimes: At long times it follows a stretched exponential which turns into a pure exponential when finite size effects dominate. As for the CTQW, the emergence of intrinsic quantum-mechanical effects, such as tunneling, prevents the decomposition of the problem into a collection of disconnected intervals and, as we will see, the mean survival probability $\Pi_M(t)$ is strongly affected by the trap arrangement. Hence, by following the temporal decay of $\Pi_M(t)$ we can extract information about the geometry. Furthermore, we show that in the cases analyzed here $P_M(t)$ and $\Pi_M(t)$ exhibit qualitatively different behaviours; this allows to determine the nature, either rather coherent or rather incoherent, of the transport process.

Our paper is structured as follows: In Sec.~\ref{sec:CTRW} we provide a brief summary of the main concepts and of the formulae concerning CTQWs. In Sec.~\ref{sec:Trapping} we introduce a mathematical formalism useful for analyzing trapping in the CTQW picture. In the following Sec.~\ref{sec:Perturbative}, we consider special arrangements of traps on a ring and we investigate the mean survival probability by means of a perturbative approach; these analytical findings are corroborated by numerical results. In Sec.~\ref{sec:random} we study the case of random distributions of traps and finally, in Sec.~\ref{sec:concl} we present our comments and conclusions.

\section{Continuous Time Quantum Walk}
\label{sec:CTRW}Let us consider a graph $\mathcal{G}$ made up of $N$ nodes and algebraically described by the so-called adjacency matrix $\mathbf{A}$:The non-diagonal elements $A_{ij}$ equal $1$ if nodes $i$ and $j$ are connected by a bond and $0$ otherwise; the diagonal elements $A_{ii}$ are $0$. From the adjacency matrix we can directly derive some interesting quantities concerning the corresponding graph. For instance the coordination number of a node $i$ is $z_i = \sum_j A_{ij}$ and the number of walks of length $\ell$ from $i$ to $j$ is given by $(A^{\ell})_{ij}$ [Biggs, 1974].

We also define the Laplacian operator $\mathbf{L}$ according to $L_{ij} = z_i \delta_{ij} - A_{ij}$; the set of all $N$ eigenvalues of $L$ is called the Laplacian spectrum of $\mathcal{G}$. Interestingly, the Laplacian spectrum is intimately related not only to dynamical processes involving particles moving on the graph, but also to dynamical processes involving the network itself; these include energy transfer and diffusion-reaction processes as well as the relaxation of polymer networks, just to name a few (see for example [Mohar, 1991; Galiceanu \& Blumen, 2007] and references therein).

In the context of coherent and incoherent transport it is worth underlining that, being symmetric and non-negative definite, $\mathbf{L}$ can generate both a probability conserving Markov process and a unitary process [Childs \& Goldstone, 2004; M\"{u}lken \& Blumen, 2005; Volta {\it et al.}, 2006].

Now, continuous-time random walks (CTRWs) [Montroll \& Weiss, 1965] are described by the following Master Equation:
\begin{equation}\label{eq:master_cl}
\frac{d}{dt} p_{k,j}(t)= \sum_{l=1}^{N} T_{kl} p_{l,j}(t),
\end{equation}
being $p_{k,j}(t)$ the conditional probability that the walker
is on node $k$ when it started from node $j$ at time $0$. If the
walk is unbiased the transmission rates $\gamma$ are bond-independent and the
transfer matrix $\mathbf{T}$ is related to the Laplacian
operator through $\mathbf{T} = - \gamma \mathbf{L}$ (in the following we set $\gamma=1$).

We now define the quantum-mechanical analog of the CTRW, i.e. the CTQW, by
identifying the Hamiltonian of the system with the classical transfer
matrix, $\mathbf{H}=-\mathbf{T}$ [Farhi \& Gutmann, 1998; M\"{u}lken \&
Blumen, 2005].
Hence, given the orthonormal basis set $|j \rangle$, representing the
walker localized at the node $j$, we can write
\begin{equation}\label{eq:tb}
\mathbf{H} = \sum_{j=1}^{N} z_j |j\rangle \langle j| -  \sum_{j=1}^{N}
\sum_{k {\rm NN} j} |k\rangle \langle j| ,
\end{equation}
where in the second sum $k$ runs over all nearest neighbors (NN) of $j$.
The operator
in Eq.~\ref{eq:tb} is also known as tight-binding Hamiltonian. Actually,
the choice of the Hamiltonian $\mathbf{H}$ is, in general, not unique
[Childs \& Goldstone, 2004] and Eq.~\ref{eq:tb} has two important
advantages: It allows to take into account the local properties of the
arbitrary substrate and, remarkably, it yields a mathematical formulation
displaying important analogies with the classical picture. In fact, the
dynamics of the CTQW can be described by the transition amplitude
$\alpha_{k,j}(t)$ from state $| j \rangle$ to state $| k \rangle$, which
obeys the following Schr\"{o}dinger equation:
\begin{equation}\label{eq:schrodinger}
\frac{d}{dt} \alpha_{k,j}(t)=-i \sum_{l=1}^{N} H_{kl} \alpha_{l,j}(t),
\end{equation}
structurally very similar to Eq.~\ref{eq:master_cl}. The solution of Eq.~\ref{eq:schrodinger} can be formally written as
\begin{equation}\label{eq:formal_solution}
\alpha_{k,j}(t)=\langle k | \exp(-i\mathbf{H}t)   | j \rangle,
\end{equation}
whose squared magnitude provides the quantum mechanical
transition probability $\pi_{k,j}(t) \equiv |\alpha_{k,j}(t)|^2$.

In general, it is convenient to introduce the orthonormal basis $|
\Phi_n \rangle, n \in [1,N]$ which diagonalizes
$\mathbf{H}$; the correspondent
set of eigenvalues is denoted by $\{E_n\}_{n=1,...,N}$. Thus, we can
write
\begin{equation}
\pi_{k,j}(t) = \left| \sum_{n=1}^{N} \langle k | e^{-i E_n t} |\Phi_n \rangle \langle \Phi_n | j \rangle \right|^2.
\end{equation}

It should be underlined that while both problems (CTRW and CTQW) are linear, and thus many results obtained in solving CTRWs (eigenvalues and eigenfunctions) can be readily reutilized for CTQWs, the physically relevant properties of the two cases differ vastly: Thus, in
the absence of traps CTQWs are symmetric under time-inversion, which precludes them from attaining equipartition for the $\pi_{k,j}(t)$ (such as the $p_{k,j}(t)$ for CTRWs) at long
times. Also, the quantal system keeps memory of the initial conditions, exemplified by the occurrence of
quasi-revivals [M\"{u}lken \& Blumen, 2005; M\"{u}lken \& Blumen, 2006].

\section{CTQWs in the presence of traps}\label{sec:Trapping}
As discussed in the previous section, the operators describing the dynamics of CTQWs and of CTRWs share the same set of eigenvalues and of eigenstates. However, when new contributions (arising e.g. from the interaction with external fields or absorbing sites) are incorporated, the eigenvalues and the eigestates start to differ. In the following we introduce a formalism useful to analyze the dynamics of CTQWs and CTRWs in the presence of traps; for this we will denote with $\mathbf{H_0}$ and $\mathbf{T_0}$ the unperturbed operators without traps.

Let us consider a system where $M$ out of the $N$ nodes are traps; we label the trap positions with $m_j$, with $j=1,...,M$, and we denote this set with $\mathcal{M}$.

For substitutional traps the system can be described by the following effective (but non-Hermitian) Hamiltonian [M\"{u}lken {\it et al.}, 2007] 
\begin{equation} \label{eq:H}
\mathbf{H} = \mathbf{H}_0 - i \mathbf{\Gamma},
\end{equation}
where $\mathbf{\Gamma}$ is the trapping operator defined as
\begin{equation} \label{eq:Gamma}
\mathbf{\Gamma} = \sum_{j=1}^M \Gamma_{m_j} | m_j \rangle \langle m_j |.
\end{equation}
The capture strength $\Gamma_{m_j}$ determines the rate of decay for a particle located at trap site $m_j$; here we will take the $\Gamma_{m_j}$ to be equal for all traps, i.e. $\Gamma_{m_j} \equiv \Gamma$ for all $j$. The limit $\Gamma \rightarrow \infty$ corresponds to perfect traps, which means that a classical particle is immediately absorbed when reaching any trap.

Due to the non-hermiticity of $\mathbf{H}$, its eigenvalues are complex and can be written as $E_l = \epsilon_l - i \gamma_l \ (l=1,...,N)$; moreover, the set of its left and right eigenvectors, $| \Phi_l \rangle$ and $\langle \tilde{\Phi}_l |$, respectively, can be chosen to be biorthonormal ($\langle \tilde{\Phi}_l |\Phi_l' \rangle = \delta_{l,l'}$) and to satisfy the completeness relation $\sum_{l=1}^N |\Phi_l \rangle \langle \tilde{\Phi}_l | = \mathbf{1}$.
Therefore, according to Eq.~\ref{eq:schrodinger}, the transition amplitude can be evaluated as
\begin{equation} \label{eq:alfa}
\alpha_{k,j}(t) = \sum_{l=1}^N e^{- (\gamma_l + i \epsilon_l)t} \langle k | \Phi_l \rangle \langle \tilde{\Phi}_l | j \rangle,
\end{equation}
from which $\pi_{k,j}(t)=|\alpha_{k,j}(t)|^2$ follows.

Of particular interest, due to its relation to experimental observables, is the mean survival probability $\Pi_M(t)$ which can be expressed as [M\"{u}lken {\it et al.}, 2007] 
\begin{eqnarray} \label{eq:pi}
\nonumber
\lefteqn{\Pi_{M} \equiv \frac{1}{N - M} \sum_{j \notin \mathcal{M} } \sum_{k \notin \mathcal{M} } \pi_{kj}(t)} \\
\nonumber
& & = \frac{1}{N-M}   \sum_{l=1}^N  e^{-2 \gamma_l t} \left ( 1- 2 \sum_{m \in \mathcal{M}} \langle \tilde{\Phi}_l | m \rangle \langle m | \Phi_l \rangle \right )  \\
& & \mbox{} + \frac{1}{N-M} \sum_{l,l'=1}^N e^{-i(E_l - E_{l'}^{*})} \left ( \sum_{m \in \mathcal{M}} \langle \tilde{\Phi}_{l'} | m \rangle \langle m | \Phi_l \rangle \right )^2 .
\end{eqnarray}
The temporal decay of $\Pi_M(t)$ is determined by the imaginary parts of $E_l$, i.e. by the $\gamma_l$.
As shown in [M\"{u}lken {\it et al.}, 2007] at intermediate and long times and for $M \ll N$ the $\Pi_M(t)$ can be approximated by a sum of exponentially decaying terms:
\begin{equation}\label{eq:pi_asym}
\Pi_{M} \approx \frac{1}{N-M} \sum_{l=1}^{N} e^{-2 \gamma_l t},
\end{equation}
and is dominated asymptotically by the smallest $\gamma_l$ values.

Now, in the incoherent, classical transport case trapping is incorporated into the CTRW according to
\begin{equation}
\mathbf{T} = \mathbf{T_0} - \mathbf{\Gamma} = - \mathbf{L} -\mathbf{\Gamma}.
\end{equation}
The transfer operator $\mathbf{T}$ is therefore real and symmetric, and it
leads to real, strictly negative eigenvalues which we denote by
$-\lambda_l$;
to them correspond the eigenstates $| \phi_l \rangle$.

Analogously, the mean survival probability for the CTRW can be written as
\begin{eqnarray}    \label{eq:p}
\nonumber
\lefteqn{P_M(t) \equiv \sum_{j \notin \mathcal{M}} \sum_{k \notin \mathcal{M}} p_{kj}(t)}   \\
& & = \frac{1}{N-M} \sum_{l=1}^{N} e^{-\lambda_l t} \left| \sum_{k \notin \mathcal{M}} \langle k | \phi_l \rangle \right|^2.
\end{eqnarray}
From Eq.~\ref{eq:p} one may deduce that $P_M(t)$ attains in general rather
quickly an exponential form; furthermore, if the smallest eigenvalue
$\lambda_{\rm min}$
is well separated from the next closest eigenvalue, $P_M(t)$ is dominated
by
$\lambda_{\rm min}$
and by the corresponding eigenstate
$|\phi_{\rm min} \rangle$
[M\"{u}lken {\it et al.}, 2007; M\"{u}lken {\it et al.}, 2008]:
\begin{equation}    \label{eq:p_asym}
P_M(t) \approx \frac{1}{N-M} e^{-\lambda_{\rm min} t} \left| \sum_{k
\notin \mathcal{M}} \langle k | \phi_{\rm min} \rangle \right|^2.
\end{equation}

Lower estimates of the gap $\Delta$ between the two smallest eigenvalues
have been found in the past for special
choices of operators (see e.g. [Chen M., 1997] and references therein).
For instance, the operator $\textbf{T}_0$ has $\lambda_{\rm min}=0$; its
next smallest eigenvalue represents the algebraic connectivity of the
graph, namely the relative number of edges needed to be deleted to
generate a bipartition. In the case of a $k$-regular graph $\Delta$ is
bounded from below by $k/(D N)$, being $D$ the diameter of the graph, i.e.
the maximum distance between any two vertices [Chung F.R.K.,
1996].

\section{Perturbative approach for trapping on a ring}\label{sec:Perturbative}
When the strength $\Gamma$ of the trap is small with respect to the couplings between neighbouring nodes (which here means $\Gamma \ll 1$), we can treat the effective Hamiltonian introduced in Eq.~\ref{eq:H} along the lines of time-independent perturbation theory.

Before developing this strategy we fix the structure $\mathcal{G}$, by considering a ring of length $N$ so that the coordination number equals $2$ for all sites ($\mathbf{Z} = 2 \mathbf{I}$), where we assume $N$ to be even.
For the corresponding Hamiltonian $\mathbf{H_0}$ we know exactly all the eigenvalues and eigenvectors; one has namely
\begin{equation}\label{eq:eigenvalues}
E_l^{(0)} = 2 - 2 \cos (2 \pi l /N)
\end{equation}
and
\begin{equation} \label{eq:eigenvectors}
| \Phi_l^{(0)} \rangle = \frac{1}{\sqrt{N}} \sum_{j=1}^{N} e^{-i 2 \pi l j /N} | j \rangle.
\end{equation}
We underline that all the eigenvalues, apart from $E_{N/2}=4$ and $E_{N}=0$, are two-fold degenerate, $E_l = E_{N-l} \, (l=1,2,...,N/2-1)$.
We now apply perturbation theory to evaluate to first order the corrections $E_l^{(1)}$ to the eigenvalues $E_l$. For $l= N/2$ and for $l=N$ we use the non-degenerate expression
\begin{equation} \label{eq:nondeg}
E_{l}^{(1)} = - i \Gamma \sum_{m \in \mathcal{M}} \left| \langle m | \Phi_l^{(0)} \rangle \right|^2
\end{equation}
and get
\begin{equation} \label{eq:correction_nondeg}
E_{N/2} = 4 - i \Gamma \frac{M}{N} \; \; \;  \mathrm{and} \; \; \; E_{N} = - i \Gamma \frac{M}{N}.
\end{equation}
For $l$ different from $N/2$ and $N$ we set
\begin{eqnarray}
V_{i,j} \equiv \langle \Phi_i^{(0)} | -i \mathbf{\Gamma} | \Phi_j^{(0)} \rangle
\end{eqnarray}
and we apply the expression valid for two-fold degenerate solutions of $\mathbf{H_0}$:
\begin{eqnarray} \label{eq:deg}
E_{l}^{(1)} = \frac{1}{2} \left ( V_{l,l} + V_{N-l,N-l} \right )   \\
\nonumber
\pm \frac{1}{2} \left[ \left( V_{l,l} - V_{N-l,N-l} \right)^2 + 4 |V_{l,N-l}|^2  \right ]^{1/2},
\end{eqnarray}
where we choose the positive sign for $l \in [1,N/2-1]$ and the negative sign for $l \in [N/2+1,N-1]$. Now we have
\begin{equation} \label{eq:V11}
V_{l,l} \equiv V_{N-l,N-l} = - i \Gamma \frac{M}{N},
\end{equation}
independently of the trap arrangement and
\begin{eqnarray} \label{eq:V12}
\nonumber
& V_{l,N-l} = -i \Gamma / N \sum_{j=1}^{M}  \exp \{ 2 i \pi m_j [l - (N-l)]/N \} \\
& = -i \Gamma / N \sum_{j=1}^{M} \exp (4 i \pi l m_j /N).
\end{eqnarray}
By inserting the last results into Eq.~\ref{eq:deg} we get
\begin{equation} \label{eq:correction_deg}
E_{l}^{(1)} = \frac{-i \Gamma}{N} \left ( M \pm  \left| \sum_{j=1}^{M}  e^{2 i \pi 2 l m_j /N} \right| \right ).
\end{equation}
We notice that for special trap arrangements the $E_l^{(1)}$ can be calculated exactly: The most striking results are obtained when the exponential in the sum in Eq.~\ref{eq:correction_deg} equals one of the values from the set $\{ 1, i, -1, -i \}$. Then the absolute value of the sum reduces to $|\sum_{j=1}^{M} \exp(i 4 \pi l m_j /N)| = M$.
For this there have to exist indices $l \neq N/2$ and $l \neq N$ such that $m_j$ can be expressed as
\begin{equation}   \label{eq:conditions} m_j = \frac{N}{8l}(4 k_j +r ) + c
\end{equation}
where $k_j$ and $c$ are arbitrary integers and $r=0,1,2$ or $3$, corresponding to $1,i,-1$ or $-i$, respectively. Consequently, we obtain for the correction
\begin{equation}
E_{l}^{(1)}= -i \Gamma \frac{M}{N} \; \; \mathrm{and} \; \;  E_{N-l}^{(1)}= 0,
\end{equation}
so that the degeneracy is always lifted.

\begin{figure}[ht]
\includegraphics[width =2.0in]{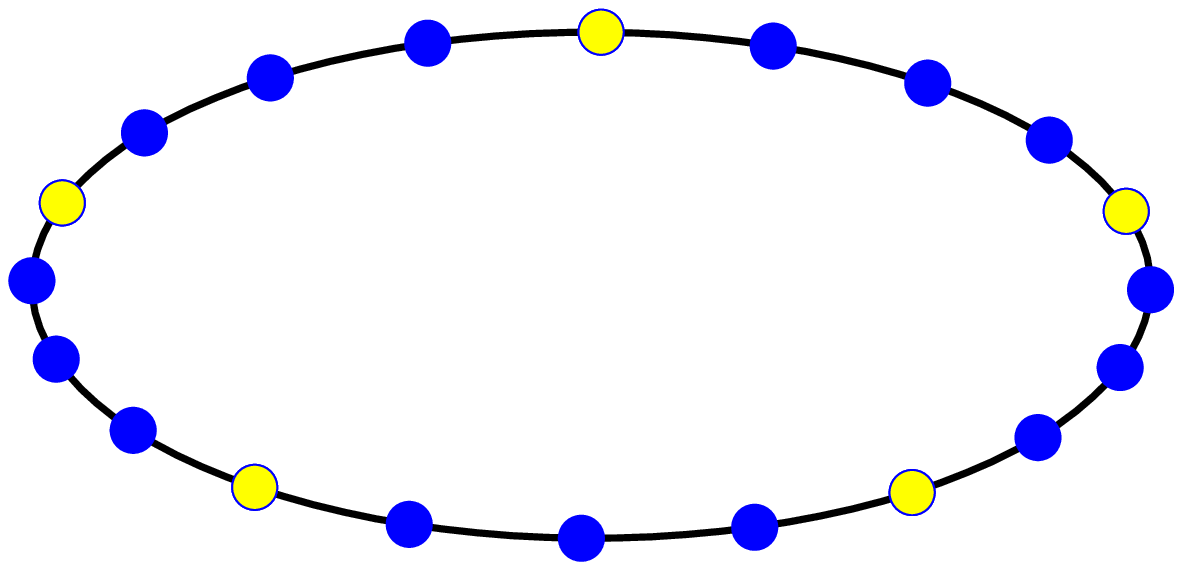}
\includegraphics[width =2.0in]{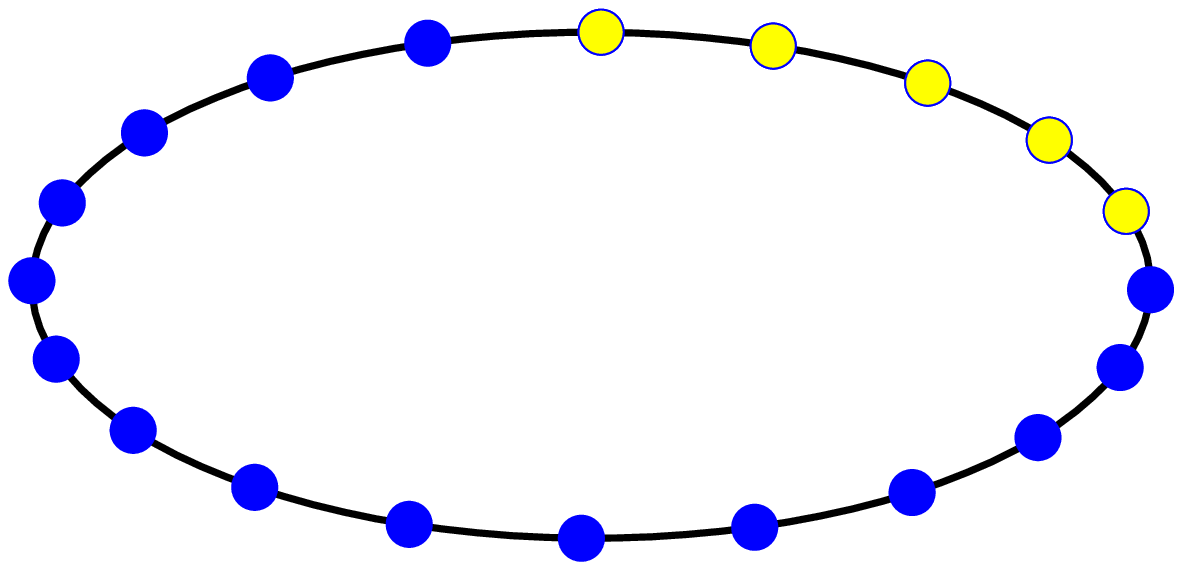}
\caption{Examples of periodic (top) and sequential (bottom) arrangements of $M=5$ traps on a ring of size $N=20$.}  \label{fig:arrangia}
\end{figure}

\begin{figure}[ht]
\includegraphics[width =3.5in]{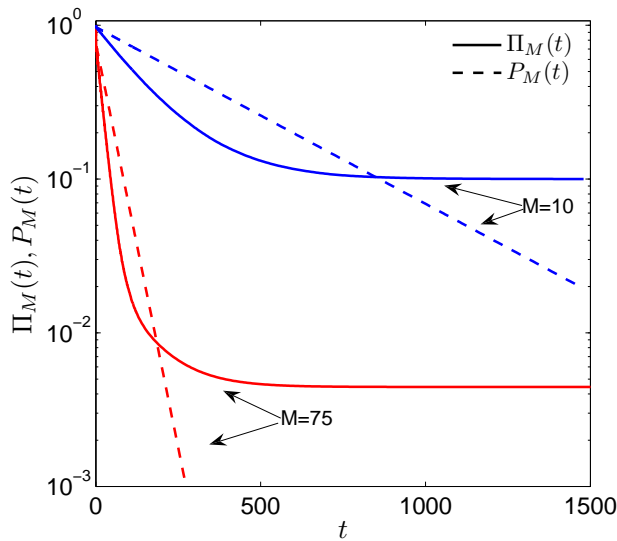}
\caption{Survival probabilities $\Pi_M(t)$ (continuous line) and $P_M(t)$ (dotted lines) on a ring of size $N=300$ and $\Gamma=0.01$ in the presence of $M=10$ and of $M=75$ traps arranged periodically, i.e. $m_j = j N/M$. Note the semilogarithmic scales.}  \label{fig:per}
\end{figure}

Let us now focus on a periodic distribution of traps with $m_j = j N/M$, while $N/M \in \mathbb{N}$. It is easy to see that in this case there exists a non-empty set $\Upsilon$ of distinct values of $l \in [1,N/2-1]$ satisfying the condition of Eq.~\ref{eq:conditions}; this occurs for $2l/M \in \mathbb{N}$, so that the cardinality of $\Upsilon$ is given by the number of integers in $\{ 2l/M \}_{l=1,2,...,N/2-1}$, namely by
\begin{equation} \label{eq:upsi}
|\Upsilon| =
\left\{
\begin{array}{cr}
\lfloor (N-2) /M \rfloor & \mathrm{\; \; for \; even} \; M, \\
\lfloor (N-2) / 2M \rfloor & \mathrm{\; \; for \; odd} \; M,
\end{array}
\right.
\end{equation}
where $\lfloor x \rfloor$ denotes the largest integer less than or equal to $x$.
In particular, for both $M=1$ and $M=2$ we have $| \Upsilon| =N/2-1$.
Now, the numerical diagonalization of the Hamiltonian $\mathbf{H}$ shows that for $l \in \Upsilon$ we get $\gamma_{N-l}=0$ (not only in first order in $\Gamma$). Consequently, the corresponding term in Eq.~\ref{eq:pi} decays to a non vanishing value, and from Eq.~\ref{eq:pi_asym} we have for $t \rightarrow \infty$:
\begin{equation} \label{eq:surv_upsilon}
\Pi_M(t) \approx \frac{|\Upsilon|}{N-M}.
\end{equation}
Hence, recalling Eq.~\ref{eq:upsi}, for large structures with $M \ll N$, $\Pi_M(t)$ asymptotically decays to $1/M$ (even case) and to $1/(2M)$ (odd case). Figure \ref{fig:per} shows results obtained for a ring of size $N=300$ with a periodic arrangement of $M=10$ ($|\Upsilon| = 29$) and $M=75$ ($|\Upsilon| = 1$) traps. Consequently, the survival probability $\Pi_M(t)$ decays to the constant values $1/10$ and $1/225$, respectively.
From a physical point of view, the finite limit for the survival probability stems from the existence of stationary states to which the nodes in $\mathcal{M}$ do not contribute, so that they never ``see" the traps. This genuine quantum-mechanical effect has no counterpart in the classical case where, for finite structures, the survival probability always decays to zero in the presence of traps. In particular, as shown in Fig.~\ref{fig:per}, $P_M(t)$ decays exponentially, as expected.

\begin{figure}[ht]
\includegraphics[width =3.5in]{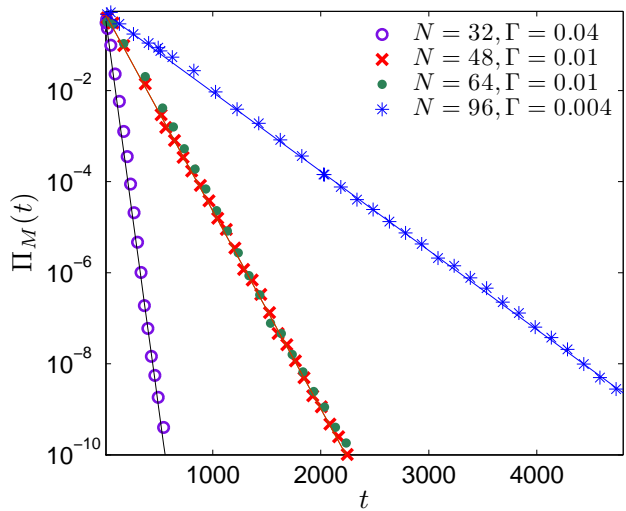}
\caption{Survival probability $\Pi_M(t)$ on rings of size $N=32, 48, 64$ and $96$ for $\Gamma=0.04, 0.01, 0.004$, as indicated. The number of traps is $M=N/2$ and they are placed consecutively, i.e. $m_j = j$. The straight lines represent Eq.~\ref{eq:pi_seq_spec}.} \label{fig:seq_q}
\end{figure}

Let us now focus on another special configuration of $M$ traps, $M > 1$: we consider a sequential arrangement, such that $m_j=j$ and $j=1,....,M$. Hence, Eq.~\ref{eq:V12} can be written as
\begin{eqnarray} \label{eq:V12_seq}
\lefteqn{V_{l,N-l} = -i \Gamma /N  \sum_{j=1}^{M} \exp (4 i \pi l j /N) }\\
\nonumber
& & = \frac{-i \Gamma}{N} \; \frac{  \exp (4 i \pi l M/N ) -1 } { \exp (4 \pi i l /N ) -1 } \; \exp (4 \pi i l /N )\\
\nonumber
& & = \frac{-i \Gamma}{N} \; \frac{\sin(2 \pi M l /N)}{\sin(2 \pi l /N)} \; \exp [2 i \pi l (M+1)/N],
\end{eqnarray}
from which we get
\begin{equation} \label{eq:correction_deg_seq}
E_l^{(1)} = \frac{-i \Gamma}{N} \left( M  \pm \frac{\sin(2 \pi M l /N)}{\sin(2 \pi l /N)} \right).
\end{equation}
We notice that since $l \neq N/2$ and $l \neq N$ then $2l/N \notin \mathbb{N}$, while for $2lM/N \in \mathbb{N}$ then $E_l^{(1)} = E_{N-l}^{(1)} = - i \Gamma M/N$. In particular, when $M=N/2$, we have $\gamma_l = M/N$ for each value of $l \in \left[ 1,N \right]$.
As a result, in Eq.~\ref{eq:pi} the first term vanishes due to the completeness property and the fact that the $\gamma_l$ are no longer $l$-dependent. As for the second term, by neglecting oscillations, we get
\begin{equation} \label{eq:pi_seq_spec}
\Pi_M (t) \approx \frac{M}{N-M} e^{-2 \Gamma t M/N} = \frac{1}{2} e^{-\Gamma t},
\end{equation}
which is independent of $N$. As shown in Fig.~\ref{fig:seq_q} the exponential behaviour predicted by Eq.~\ref{eq:pi_seq_spec} holds also for intermediate times.

\begin{figure}[ht]
\includegraphics[width =3.5in]{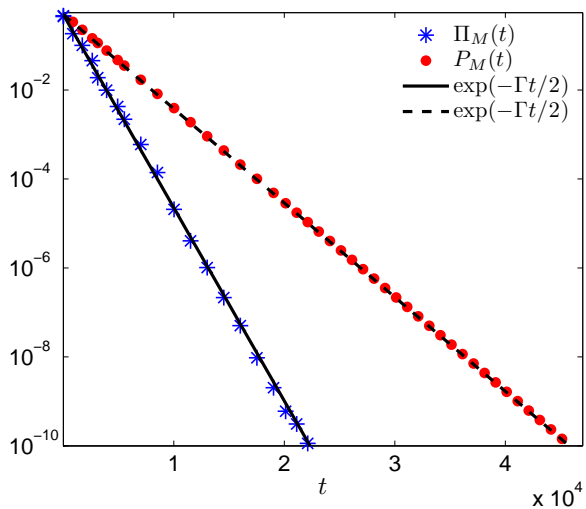}
\caption{Survival probabilities $\Pi_M(t)$ and $P_M(t)$ on rings of size $N=48$ for $\Gamma=0.001$; the number of traps is $M=N/2$ and they are placed consecutively, i.e. $m_j = j$.
The straight lines represent Eq.~\ref{eq:pi_seq_spec} (continuous line) and Eq.~\ref{eq:p_seq_spec} (dashed line).}  \label{fig:seq_cq}
\end{figure}

In Fig.~\ref{fig:seq_cq} we compare the survival probabilities of CTQWs and CTRWs: as highlighted by the semi-logarithmic plot, the decay is exponential in both cases, although faster in the former. Indeed, for the CTRWs we have in the long-time limit from Eq.~\ref{eq:p}:
\begin{equation}\label{eq:p_seq_spec}
P_M(t)\approx \frac{N-M}{N}  e^{- \Gamma M t /N} = \frac{1}{2} e^{- \Gamma t /2},
\end{equation}
where we used the fact that the smallest eigenvalue is $\Gamma M /N$. By comparing Eq.~\ref{eq:pi_seq_spec} and Eq.~\ref{eq:p_seq_spec} we see that, although the decay is exponential in both cases, the decay rate is twice larger for $\Pi_M(t)$ than for $P_M(t)$.

\section{Random distributions of traps}\label{sec:random}
\begin{figure}[ht]
\includegraphics[width =3.5in]{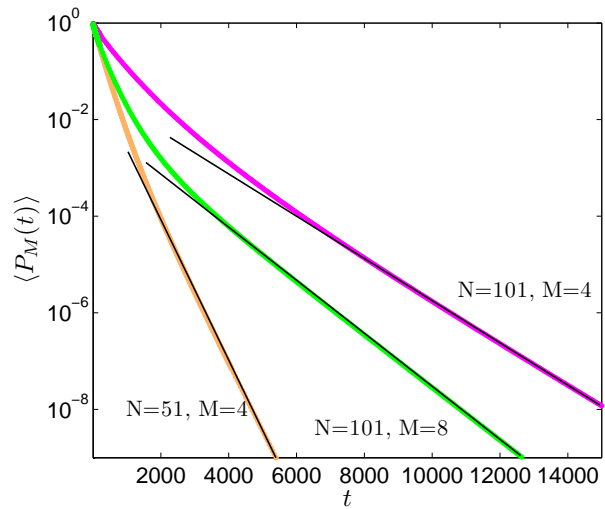}
\caption{Average survival probabilities $\langle P_M(t) \rangle$ for rings of sizes $N=51$ and $101$. Here $\Gamma=0.1$ and $M$ is either $4$ or $8$. The data presented have been averaged over $120$ different realizations, see text. The straight lines highlight the exponential decay.}  \label{fig:rand_c}
\end{figure}

We now take $N$ to be odd (so as not to fulfill Eq.~\ref{eq:conditions}) and consider random arrangements of traps: we pick $M$ distinct trap locations randomly from a uniform distribution and determine the corresponding $\Pi_M(t)$ and $P_M(t)$. Then we average these over different, independent realizations to determine $\langle \Pi_M(t) \rangle$ and $\langle P_M(t) \rangle$. As already mentioned in Sec.~\ref{sec:intro}, $\langle P_M(t) \rangle$ exhibits different behaviours: in an infinite system the decay law is a stretched exponential at long times, whereas in finite systems at such times the decay gets to be exponential. In Fig.~\ref{fig:rand_c} we show evidence of the long-time exponential behaviour of $\langle P_M(t) \rangle$ in systems of relatively small size.

\begin{figure*}[ht]
\includegraphics[width =4.2in]{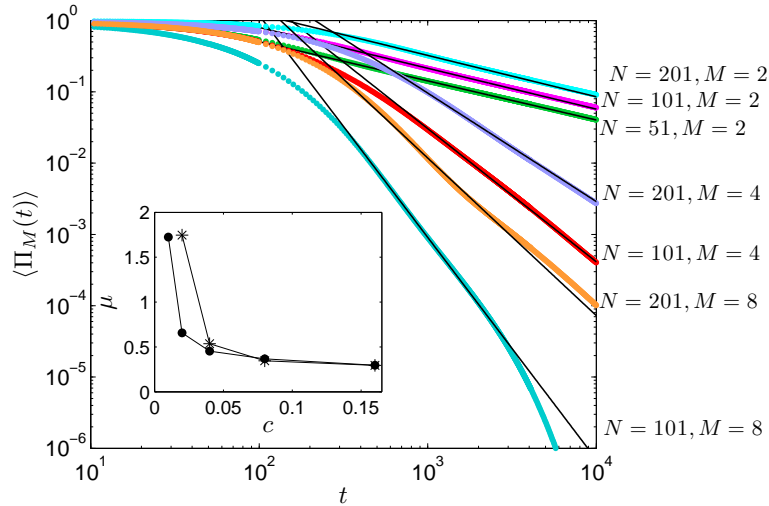}
\caption{$\langle \Pi_M(t) \rangle$ and $\mu$ for rings of sizes $N=51, 101,$ and $201$. Here $\Gamma=0.1$ and $M$ is $2,4,$ or $8$, as indicated. The data presented have been averaged over $120$ different realizations.
The main figure displays the average survival probabilities $\langle \Pi_M(t) \rangle$ in double logarithmic scales. The straight lines represent the best fit. The inset displays the exponent $\mu$ as a function of $c=M/N$ for systems of sizes $N=201$ ($*$) and $N=101$ ($\bullet$).}  \label{fig:rand}
\end{figure*}

Let us now consider $\langle \Pi_M(t) \rangle$ for random trap arrangements. Now, the $\langle \Pi_M(t) \rangle$ decay differs qualitatively from that of the $\Pi_M(t)$ analyzed in the previous section. As shown in Fig.~\ref{fig:rand}, for intermediate times the average survival probability displays a power law, which decays more slowly than exponentially:
\begin{equation}
\langle \Pi_M(t) \rangle \sim t^{-1 / \mu}.
\end{equation}
A similar result has already been obtained for CTQWs on a finite chain with two traps at its ends ($\mathcal{M}=\{1,N\}$), in the presence of either nearest-neighbour [M\"{u}lken {\it et al.}, 2007]
or long-range interactions [M\"{u}lken {\it et al.}, 2008]. There one could understand the power law decay based on the imaginary part of the Hamiltonian spectrum $\{ \gamma_l \}$, which in a large interval scales algebraically with $l$.
By fitting the numerical data obtained for different sizes and concentrations we get the characteristic exponent $\mu$ depicted in the inset of Fig.~\ref{fig:rand}.

\section{Conclusions}\label{sec:concl}
In conclusion, we have modeled the coherent dynamics by continuous-time quantum walks and studied interactions with traps: Taking a periodic chain as substrate, we calculated the mean quantal survival probability $\Pi_M(t)$ and we compared it to the classical $P_M(t)$ for different trap arrangements. The quantum problem was approached both analytically (by means of perturbative theory) and numerically, showing that the spatial distribution of the traps significantly affects $\Pi_M(t)$.
In particular, when the traps are arranged periodically throughout the substrate, $\Pi_M(t)$ decays asymptotically to a nonvanishing value which depends directly on the system size $N$ and on the number of traps $M$ (e.g., when $M=2$, $\Pi_2(t) \rightarrow 1/2$ for $t \rightarrow \infty$). This is a genuine quantum-mechanical effect with no counterpart in classical mechanics, where $P_M(t)$ decays to zero for finite systems.

Another interesting, deterministic trap configuration is realized by distributing the traps consecutively such to form a cluster; then at intermediate and long times the survival probability decays exponentially with the characteristic time $\Gamma^{-1}$. Now, for the same trap configuration, the characteristic time for the classical survival probability doubles, being $2 \Gamma^{-1}$.

When the traps are distributed randomly on the substrate, a further, qualitatively different behaviour of $\Pi_M(t)$ is obtained. In fact, by averaging over different independent configurations we find in this case that at intermediate times $\langle \Pi_M(t) \rangle$ decays algebraically, i.e. $\langle \Pi_M(t) \rangle \sim t^{-1 / \mu}$, where $\mu$ depends on $M$ and $N$ and is related to the imaginary part of the Hamiltonian spectrum. On the other hand, for systems of relatively small size we find that in the same time range finite-size effects dominate $\langle P_M(t) \rangle$, giving rise to an exponential decay.

These results establish that studying the decay due to trapping is indeed an advantageous means to monitor the system's evolution, as it allows to determine the nature of the transport, which can be either rather coherent or rather incoherent. Moreover, the behaviour exhibited by $\Pi_M(t)$, being qualitatively affected by the trap configurations, may be used to distinguish between these.

\bigskip

\noindent {\bf Acknowledgments} \smallskip Support from the Deutsche Forschungsgemeinschaft (DFG) and the Fonds
der Chemischen Industrie is gratefully acknowledged.
EA thanks the Italian Foundation ``Angelo Della Riccia" for financial support.

\bigskip

\noindent {\bf References} \smallskip

\noindent Agliari, E., Blumen, A. \& M\"{u}lken, O. [2008] ``Dynamics of
Continuous-time quantum walks in restricted geometries," {\it J. Phys. A} {\bf 41}, 445301-445321.

\noindent Anderson, P.W. [1958] ``Absence of Diffusion in Certain Random Lattices," {\it Phys. Rev.} {\bf 109}, 1492-1505.

\noindent ben-Avraham D. \& Havlin S. [2000] ``Diffusion and Reactions in Fractals and Disordered Systems," Cambridge University Press.

\noindent Biggs N. [1974] ``Algebraic graph theory," Cambridge University Press.

\noindent Blumen A., Klafter, J. \& Zumofen G. [1983] ``Trapping and reaction rates on fractals," {\it Phys. Rev. B} {\bf 28}, 6112-6115.

\noindent Blumen A., Klafter, J. \& Zumofen G. [1986] ``Models for reaction dynamics in glasses," in {\it Optical Spectroscopy of Glasses}, I. Zschokke ed.,  D. Reidel, Dordrecht , pp. 199-265.

\noindent Chen M. [1997] ``Coupling, spectral gap and realted topics (II)," {\it Chinese Science Bulletin} {\bf 42}, 1409-1416.

\noindent Childs A.M. \& Goldstone J. [2004] ``Spatial search by quantum walk," {\it Phys. Rev. A} {\bf 70}, 022314-022324.

\noindent Chung F.R.K. [1996] ``Spectral graph theory," CBMS Lecture Notes.

\noindent Farhi E. \& Gutmann S. [1998] ``Quantum computation and decision trees," {\it Phys. Rev. A} {\bf 58} 915-928.

\noindent Galiceanu M. \& Blumen A. [2007] ``Spectra of Husimi cacti: Exact results and applications," {\it J. Chem. Phys.} {\bf 127}, 134904-134911.

\noindent Hemenger R.P., Lakatos-Lindenberg K. \& Pearlstein R.M. [1974]
``Impurity quenching of molcular excitons. III. Partially coherent excitons in linear chains," {\it J. Chem. Phys.} {\bf 60}, 3271-3277.

\noindent Huber, D.L. [1980]
``Fluorescence in the presence of traps. II. Coherent transfer," {\it Phys. Rev. B} {\bf 22}, 1714-1721.

\noindent Kempe, J. [2003]
``Quantum random walks: an introductory overview," {\it Contemp. Phys.} {\bf 44}, 307-327.

\noindent Kenkre, V.M. [1978]
``Model for Trapping Rates for Sensitized Fluorescence in Molecular Crystals," {\it Phys. Status Solidi B} {\bf 89}, 651-654.

\noindent Mohar B., [1991]
``The Laplacian Spectrum of Graphs''. In
{\it Graph Theory, Combinatorics, and Applications}, Vol. 2, Ed. Y. Alavi,
G. Chartrand, O.R. Oellermann, A.J. Schwenk, Wiley, pp.\ 871-898.

\noindent Montroll E.W. \& Weiss G.H. [1965]
``Random walks on lattices II," {\it J. Math. Phys.} {\bf 6} 167-181.

\noindent M\"{u}lken O., Bierbaum V. \&  Blumen A. [2006] ``Coherent exciton transport in dendrimers and continuous-time quantum walks," {\it J. Chem. Phys.}
{\bf 124}, 124905-124911.

\noindent M\"{u}lken O. \&  Blumen A. [2005] ``Slow transport by continuous-time quantum walks," {\it Phys. Rev. E}
{\bf 71}, 016101-016106.

\noindent M\"{u}lken O. \&  Blumen A. [2006] ``Continuous-time quantum walks in phase space," {\it Phys. Rev. A} {\bf 73}, 012105-012110.

\noindent M\"{u}lken O., Blumen A., Amthor T., Giese C., Reetz-Lamour M. \& Weidem\"{u}ller M. [2007] ``Survival Probabilities in Coherent Exciton Transfer with Trapping," {\it Phys. Rev. Lett.} {\bf 99}, 090601-090605.

\noindent M\"{u}lken O., Pernice V. \& Blumen A. [2008] ``Slow Excitation Trapping in Quantum Transport with Long-Range Interactions," {\it Phys. Rev. E} {\bf 78} 021115-021119.

\noindent Olaya-Castri A., Lee C.F., Fassioli Olsen F. \& Johnson N.F. [2008] ``Efficiency od energy transfer in a light-harvesting system under quantum coherence," {\it Phys. Rev. B} {\bf 78} 085115-085121.

\noindent Parris, P.E. [1991]
``Quantum and Stochastic Aspects of Low-Temperature Trapping and Reaction Dynamics," {\it J. Stat. Phys.} {\bf 65}, 1161-1172.

\noindent Sillanp\"{a}\"{a} M.A., Park J.I. \& Simmonds R.W. [2007] ``Coherent quantum state storage and transfer between two phase qubits via a resonant cavity," {\it Nature} {\bf 449}, 438-442.

\noindent Van Kampen N.G. [1981] ``Stochastic Processes in Physics and Chemistry," North-Holland, Amsterdam.

\noindent Volta A., M\"{u}lken O. \& Blumen A. [2006] ``Quantum transport on two-dimensional regular graphs," {\it J. Phys. A} {\bf 39}, 14997-15012.

\noindent Woerner M., Reimann K. \& Elsaesser T. [2004] ``Coherent charge transport in semiconductor quantum cascade structures," {\it J. Phys.: Condens. Matter} {\bf 16}, R25-R48.

\noindent Zhou L., Gong Z.R., Liu Y.-X., Sun C.P. \& Nori F. [2008] ``Controllable Scattering of a Single Photon inside a One-Dimensional Resonator Waveguide," {\it Phys. Rev. Lett.} {\bf 101}, 100501-100505.

\end{document}